# Reconfigurable wavelength-encoded stochastic illumination for active hyperspectral imaging

Yi-Jing Chen[1], Bao-Lei Liu[1,2,*], Ze-Yuan Dong[1], Zhi-Hao Zhao[1], Yi-Ying Zhang[1] , Chun-Min Yu[1], Zhi-Hua Xu[3], Yuan-Jin Yu[4,*], Zhao-Hua Yang[1,2,*]

[1] School of Instrumentation Science and Optoelectronic Engineering, Beihang University, Beijing 100191, China

[2] Hangzhou International Innovation Institute, Beihang University, Hangzhou, 311115, China

[3] Beihang University Hospital, Beihang University, Beijing 100191, China

[4] School of Automation, Beijing Institute of Technology, Beijing 100081, China

[*] Correspondence: liubaolei@buaa.edu.cn; yuanjin.yu@bit.edu.cn; yangzh@buaa.edu.cn

**Abstract**

Traditional hyperspectral imaging (HSI) relies on sequential scanning with complex and bulky hardware, inherently limiting its temporal resolution while increasing system complexity and cost. Computational HSI offers cost-effective alternatives with simplified hardware. However, most existing computational methods rely on fixed spectral encoding units, which lack adaptability for different spectral tasks. Here, we present a reconfigurable optical stochastic encoding (ROSE) framework with programmable illumination, which can be adaptively optimized for different spectral tasks, for high-throughput, compressive HSI. By leveraging an array of monochromatic light-emitting diodes (LEDs), we synthesize stochastic spectral patterns that enable compressive acquisition using a standard monochrome camera. The proposed framework allows dynamic reconfiguration of illumination patterns, making it adaptable to diverse imaging requirements. We experimentally validate the proposed method and achieve HSI with a spatial resolution of 2048 × 1536, reconstructing 60 spectral bands across the spectral range of 400–700 nm. Furthermore, we introduce an automatic optimization strategy to search for optimal illuminations tailored to specific tasks, improving both reconstruction accuracy and task-oriented performance. We demonstrate the effectiveness of our approach in applications including anti-counterfeiting inspection and oral imaging, and further validate its compatibility with standard microscope and endoscope systems. The developed ROSE illumination module could serve as a universal, plug-and-play add-on for conventional cameras and existing optical systems, providing a cost-effective pathway to upgrade them into high-performance, task-adaptive HSI systems.

## Introduction

Hyperspectral imaging (HSI), which simultaneously captures spatial and spectral information, has attracted increasing attention in recent years due to its broad applicability in precision agriculture (*1*), early disease screening (*2*), and remote sensing (*3*). By acquiring reflectance or radiance information over contiguous spectral bands, HSI provides substantially richer material discrimination and characterization capabilities than conventional RGB imaging. However, traditional HSI methods rely on mechanical scanning in spectral or spatial dimension (*4, 5*), which severely limits their practical deployment in both the acquisition time and the hardware complexity (*6*).

To overcome these limitations, computational HSI methods have been investigated. Leveraging compressed sensing (CS) or data-driven deep learning strategies, a variety of compressive HSI methods have been proposed (*7, 8*), such as the coded aperture snapshot spectral imaging (CASSI) (*9*) and compact snapshot HSI (*10*). By combining dispersive elements with coded apertures, CASSI modulates spectral information across different wavelength bands, allowing a detector to capture spectrally multiplexed measurements in a single exposure, followed by spectral reconstruction using CS algorithms (*11*). Nevertheless, the introduction of multiple optical components and spectral modulation stages makes CASSI systems highly susceptible to system noise and model mismatch, which degrades reconstruction accuracy and robustness (*12, 13*).

The compact snapshot HSI has emerged as a promising solution due to its significantly simplified optical architecture (*14*). In these approaches, the spectral cube is spatial-spectral encoded into two-dimensional measurements by directly employing spectral-encoded filter devices (*15, 16*), such as tailored multispectral filter arrays (*17-19*), quantum-dot-based structures (*20-22*), and plasmonic metasurfaces (*23-25*). However, such custom-designed filters usually need complex fabrication processes and cannot be further tuned once fabricated, which leads to poor flexibility to meet task-specific demands in practical applications (*26-28*). Meanwhile, most computational

HSI methods rely on passive spectral encoding implemented along the optical detection path, which is limited by low light throughput after spectral filtering and reduced spatial resolution (*29*). In contrast, the active encoded illumination methods have emerged with high light throughput and high compatibility to existing systems, but also rely on fixed spectral filters or complex electroluminescent devices (*30, 31*).

In this work, we present a reconfigurable optical stochastic encoding (ROSE) framework for active and compressive HSI, which can be adaptively optimized for different spectral tasks. By leveraging an array of monochromatic light-emitting diodes (LEDs), we synthesize stochastic spectral illumination. Furthermore, a genetic algorithm (GA) is introduced to optimize the illumination encoding strategy, allowing enhanced spectral resolution and reconstruction accuracy within selected wavelength ranges of interest, offering superior adaptability and optimization capability for compressed HSI. As an add-on illumination module to existing optical systems, ROSE supports achieving hyperspectral images with a high spatial resolution of 2048 × 1536, without the need to sacrifice spatial resolution in typical snapshot HSI methods. We validate the effectiveness of ROSE framework across diverse targets and applications, and further demonstrate its compatibility with standard optical systems, enabling their upgrade to HSI.

## PRINCIPLE

The schematic of the ROSE-based HSI system is illustrated in Fig. 1. As shown in Fig. 1(a), the system consists of an encoded illumination module composed of multiple narrowband LEDs with different central wavelengths driven by a pulse-width modulation (PWM) circuit for intensity modulation, and a CMOS camera equipped with an imaging lens. During the acquisition of raw images under the encoded illuminations, different combinations of LED sources are sequentially activated according to predefined wavelength encoding strategies, and the encoded illumination is projected onto the surface of the object. Let $S_{ref}(x, y, \lambda)$ denote the spectral response of the object at spatial position $(x, y)$, and let $D(\lambda)$ represent the spectral response function of the CMOS camera. Under a specific wavelength-encoded illumination $R_i(\lambda)$, the CMOS camera performs a single exposure and acquires a corresponding raw image $I_i(x, y)$. The raw image $I_i(x, y)$ represents the integrated response of the object in the full spectral bands under the wavelength-encoded stochastic illumination, as:

$$I_i(x,y) = \int S_{ref}\left(x,y,\lambda\right)\cdot D\left(\lambda\right)\cdot R_i\left(\lambda\right)d\lambda\,. \tag{1}$$

As shown in Fig. 1(b), the wavelength-encoded stochastic illumination can be flexibly designed through two degrees of freedom: the combination of selected LEDs and the emission intensity of each LED. Accordingly, the spectra of final illumination patterns can be regarded as a superposition of LEDs with different wavelengths under different modulation weights, and the spectrum of each encoded stochastic illumination can be denoted as:

$$R_i(\lambda) = \sum_{j=1}^{J} b_{i,j}\cdot w_{i,j}\cdot L_j(\lambda), \tag{2}$$

where $L_j(\lambda)$ represents the intrinsic spectral response of the $j$-th LED, $J$ represents number of distinct narrowband LEDs, $b_{i,j} \in \{0,1\}$ is a binary decision variable indicating whether the $j$-th LED is selected for the $i$-th illumination, and $w_{i,j} \in [0,1]$ denotes the corresponding PWM duty cycle weight to modulate the emission intensity.

Figure 1(c) presents the reconstruction process of hyperspectral images. Under different wavelength-encoded illumination, a total of $n$ two-dimensional raw images are sequentially captured. The hyperspectral image cube $S_{rec}$ consisting of $N$ spectral bands ($N \gg n$) can be reconstructed by solving the:

$$S_{rec}(x,y,\lambda) = \arg\min_{S_{rec}} \sum_{i=1}^{n} \left\| I_i(x,y) - \int S_{rec}\left(\lambda\right)D\left(\lambda\right)R_i\left(\lambda\right)d\lambda \right\|_2^2. \tag{3}$$

The objective of the reconstruction is to find an optimal $S_{rec}$ that minimizes the discrepancy between the measurements $I_i(x,y)$ and the predicted values derived from the forward physical model. This reconstruction process can be implemented using compressive sensing algorithms or data-driven methods (*32-34*) (see Supplementary Note 6 for more details about HSI reconstruction method).

With a limited number of exposure raw frames, fixed wavelength encoding schemes cannot achieve optimal reconstruction performance across different spectral bands for different imaging task requirements. To flexibly accommodate different application requirements in terms of both spectral resolution and reconstruction accuracy, we propose a genetic-algorithm (GA)-based wavelength encoding optimization framework that automatically searches for adaptive encoding strategies (*35-37*). The optimization objective is defined as the root mean square error (RMSE) evaluated across $N$ reconstructed spectral bands within a specific wavelength range of interest $\Phi_{target}$. Thus, the GA aims to identify optimal encoding strategies $(b,w)$ by minimizing the

fitness function:

$$E(b,w)=\sqrt{\frac{1}{N}\sum_{\lambda_n\in\phi_{target}}\left(S_{ref}(\lambda_n)-S_{rec}(\lambda_n,b,w)\right)^2}\ , \tag{4}$$

where $\lambda_n(n=1,2,\dots,N)$ denotes the sampling points within the targeted range $\Phi_{target}$ determined by the total number of reconstructed bands $N$. $S_{rec}(\lambda_n,b,w)$ denotes the reconstructed spectrum obtained through the active illumination encoding. For minimizing $E(b,w)$, the GA optimize the illumination coding matrix specifically for the designated spectral band, ensuring high-fidelity reconstruction where it is most needed. More detailed process is described in Supplementary Note 4. As illustrated in Figure 1(d), the ROSE enables fine-grained spectral sampling for distinct tasks, such as broadband reconstruction across the 400–700 nm range (top) or targeted narrowband reconstruction within the 600–700 nm region (bottom).

Meanwhile, serving as a plug-and-play illumination module, ROSE could be integrated with a monochrome CMOS camera. The system acquires 16 encoded raw images under stochastic illumination within 0.5 seconds, and then reconstructs 60 spectral bands across the 400–700 nm range with a spatial resolution of 2048 × 1536 pixels, without sacrificing spatial resolution of the CMOS camera.

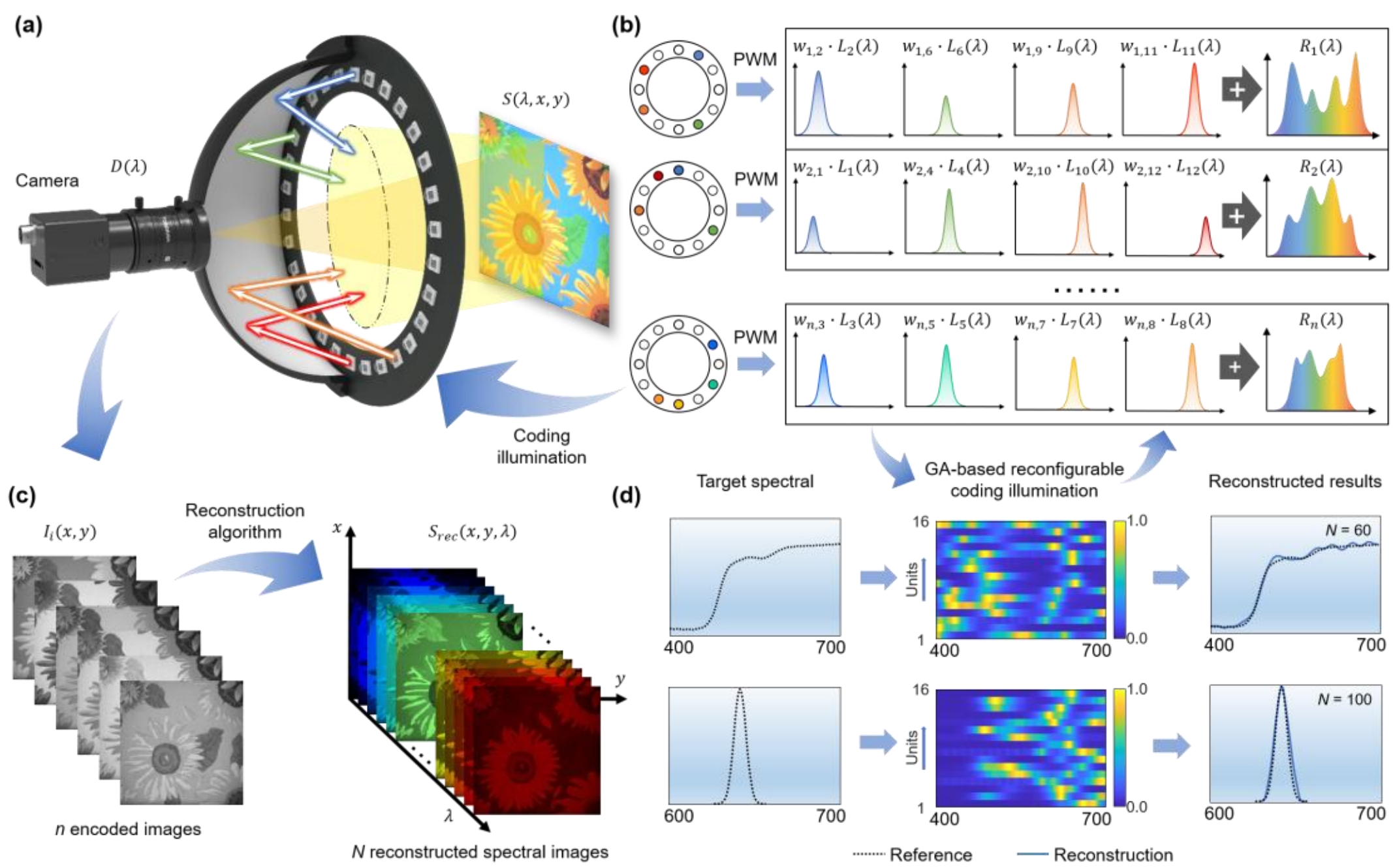

**Figure 1**. Schematic and working principle of the proposed reconfigurable optical stochastic encoding (ROSE) illumination for HSI. (a) Schematic of ROSE-based HSI, which consists of an encoded illumination module composed of multiple narrowband LEDs, and a standard monochrome camera. (b) Synthesis of illumination patterns with stochastic spectra by driving diverse LED subsets via pulse-width modulation (PWM). (c) Reconstruction of hyperspectral images with the input of undersampled raw images. (d) Illumination patterns of ROSE can be reconfigured and optimized for tailored spectral bands, enabling fine-grained spectral sampling for distinct reconstruction tasks.

## RESULTS AND DISCUSSION

### A. System Configuration

The structural design and encoding process of ROSE illumination module are illustrated in Figure 2. The illumination module consists of 30 monochromatic LEDs with distinct center wavelengths, whose normalized spectral response curves are shown in Figure 2(a). The center wavelengths of the LEDs span the range from 380 nm to 750 nm, with full width at half maximum (FWHM) values ranging from 14 nm to 40 nm, enabling spectral modulation across the entire wavelength range (see Supplementary Note 1 for details). Figure 2(b) illustrates the wavelength encoding process obtained by the combination of any two LEDs with different center wavelengths, in which each PWM control channel supports 256 discrete intensity levels. This flexibility in combining LEDs with different wavelengths and intensities enables the optimization of illumination patterns for diverse spectral imaging tasks.

To further improve the spatial and spectral uniformity of the illumination, the module is designed within a diffuse dome structure, as shown in Figure 2(c). Four LEDs with identical center wavelengths are grouped into a sub-array and uniformly distributed in a circular layout on the circuit board, resulting in a total of 120 narrowband LEDs (see Supplementary Note 1 for details). This layout enables multi-angle illumination of the target and alleviates shadow artifacts caused by unilateral illumination. To ensure spectral consistency of illumination across different spatial locations, a dome structure is incorporated into ROSE illumination module. Light emitted by the LEDs undergoes multiple reflections on the inner surface of the dome before illuminating the object (see Supplementary Note 3 for details about the structural design). As shown in Figure 2(c), the spectral curves at Position 1 and Position 2 demonstrate high consistency between the central and surrounding regions. The amplitude differences can be eliminated after calibration.

The prototype of ROSE illumination module is shown in Figure 2(d). The device has a compact cubic form factor with dimensions of $20 \times 22 \times 17$ $cm^3$, and the illumination aperture is a circular region with a diameter of 13.6 cm. The module can synthesize diverse stochastic illumination spectra, as showcased by the representative spectral curves in Figure 2(e). Figure 2(f) presents the correlation matrix of these illumination patterns. This low correlation ensures linear independence of the measurements, which is favorable for the spectral reconstruction.

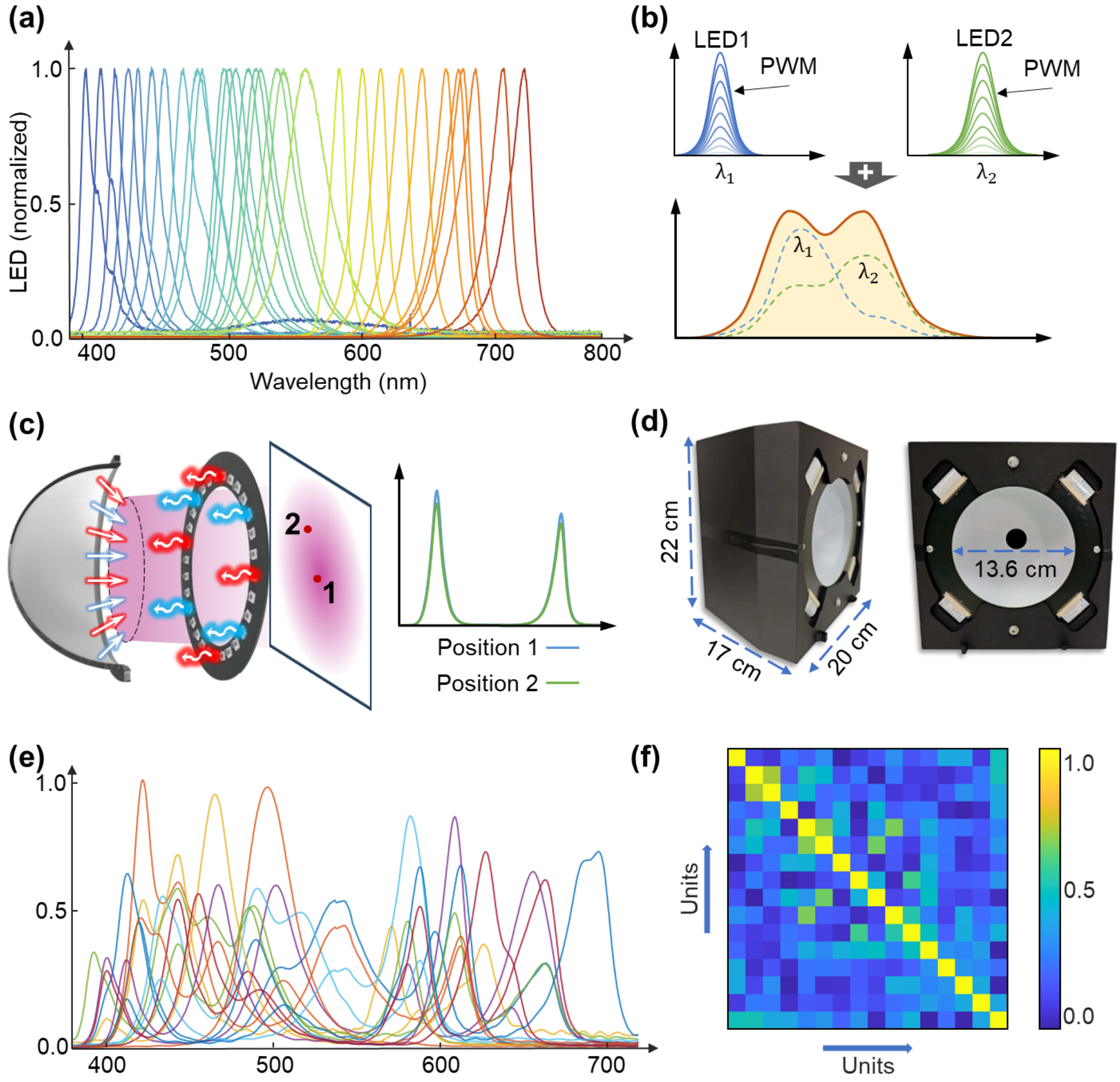

**Figure 2.** Structural design and encoding process of ROSE illumination module. (a) Normalized emission spectral curves of the 30 narrowband LEDs with different center wavelengths. (b) Illustration of the encoding process achieved by the combination of two LEDs with different center wavelengths. (c) Illumination uniformity enhancement using a centrally symmetric ring arrangement and a diffuse dome structure. (d) The prototype of ROSE illumination module. (e) Representative spectral curves of the synthesized stochastic illumination patterns. (f) Correlation matrix of the encoded illumination patterns.

## B. Validation of the reconstructed spectra for different tasks

To validate the adaptive encoding capabilities of the ROSE framework across diverse sensing tasks, spectral reconstruction experiments are conducted using a broadband target spanning the 400–700 nm range, alongside narrowband targets situated in the 400–500 nm, 500–600 nm, and 600–700 nm regions.

A color chart (X-Rite ColorChecker Classic, 24-color chart) is utilized as the target for broadband spectral reconstruction, as shown in Figure 3(a). Figure 3(b) presents the spectra of the encoded illumination patterns that were optimized for the wide-band spectral reconstruction task, which comprises 16 encoded illumination patterns covering the continuous 400–700 nm range (more details can be found in Supplementary Note 5). The reconstruction spectra, corresponding to the four selected areas, are presented in Figure 3(c). The reconstructed spectra for the four selected color patches show good agreement with the reference spectra in both overall trends and fine spectral structures, reconstructing 60 spectral bands with an average root-mean-square error (RMSE) of 0.024 (more results can be found in Supplementary Note 7).

To further evaluate the ROSE framework for high-spectral-resolution tasks, narrowband optical filters with center wavelengths of 450 nm, 532 nm, and 635 nm with the FWHM of 10 nm are used as transmission targets. The comparative results before and after task-constrained optimization are presented in Figure 3(d). When the same illumination patterns from Figure 3(b) that were previously optimized for the wide-band task are applied directly, the reconstruction accuracy degrades significantly, leading to inaccurate recovery of the filters' narrowband transmission characteristics, as shown in the left column of Figure 3(d).

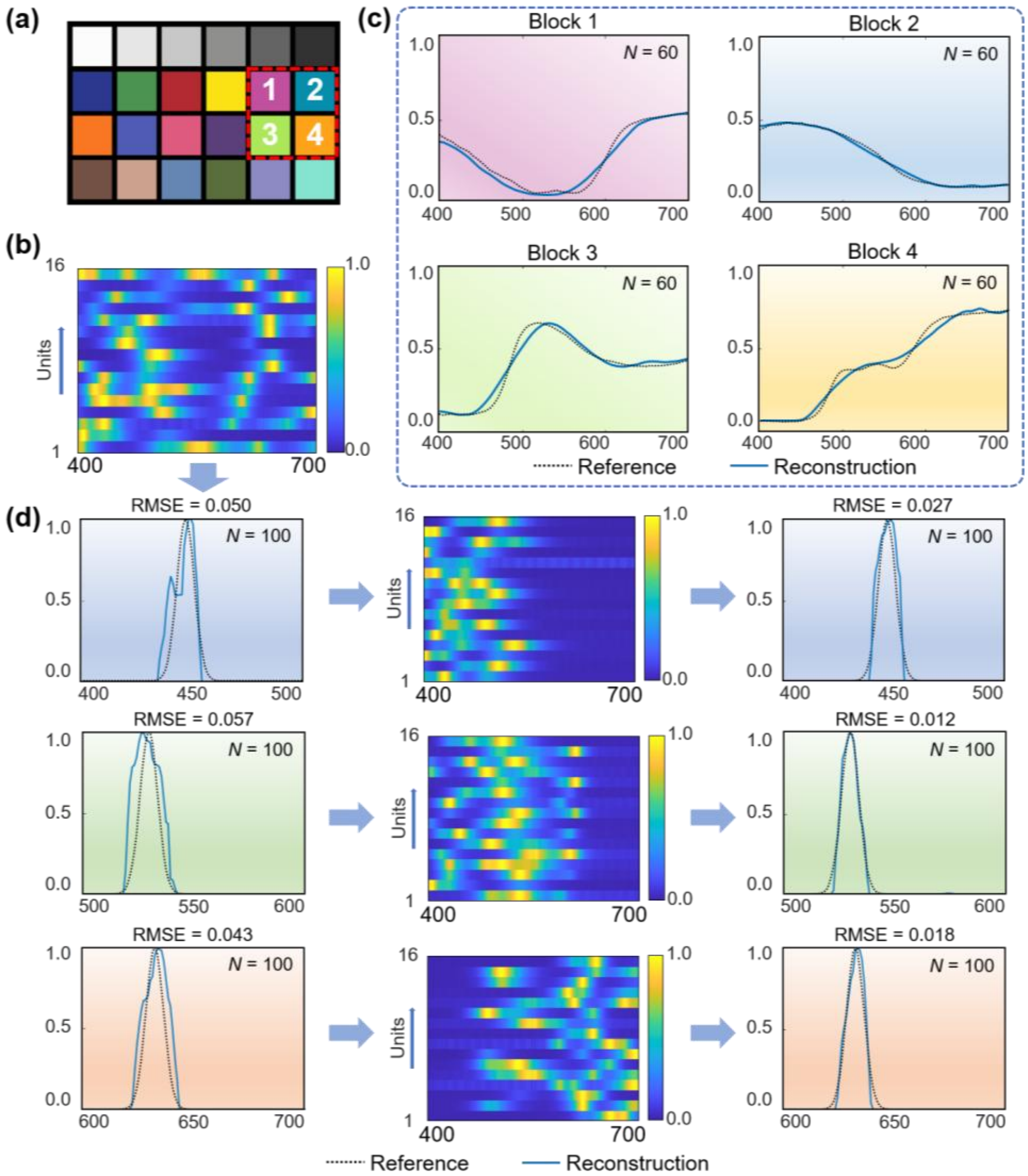

**Figure 3.** Reconfigurable wavelength encoding for spectral reconstruction. (a) The 24-color chart used as the target for broadband spectral acquisition. (b) The encoding matrix optimized for broadband (400–700 nm) spectral reconstruction. (c) Comparison of the reconstructed and reference spectra for the four selected color patches under the wide-band encoding strategy. (d) Comparative spectral reconstruction results for narrowband optical filters before and after task-constrained optimization.

To address this, a task-constrained wavelength encoding optimization strategy is introduced. Specifically, separate optimization objectives are defined for the localized ranges of 400–500 nm, 500–600 nm, and 600–700 nm, respectively. This guides the system to adaptively adjust the encoding matrix to concentrate the stochastic spectral encoding within the respective bands of interest, as presented in the middle column of Figure 3(d), maintaining the number of 16 illumination patterns. As demonstrated in the right column of Figure 3(d), this task-specific optimization significantly improves the reconstructed spectral peak positions, bandwidths, and spectral shapes, enabling an enhanced reconstruction of 100 spectral bands. Concurrently, the average RMSE is drastically reduced from 0.050 to 0.019. These results clearly highlight the effectiveness and flexibility of the ROSE framework in adaptively adjusting its encoding strategy to meet the requirements of diverse, high-precision spectral sensing tasks.

**C. Hyperspectral Imaging of the representative reflective object**

To further evaluate ROSE, HSI experiments were conducted on a reflective object, as shown in Figure 4. A cartoon-style oil painting was selected as the target and placed at the output plane of the illumination module during data acquisition. Owing to its rich textures and diverse color regions, the sample enables a comprehensive evaluation of the system's reconstruction performance. The illumination patterns used in Figure 4 were optimized for the 400-700 nm spectral band. Figure 4(b) presents 5 representative wavelengths selected from the reconstructed hyperspectral cube (more results can be found in Supplementary Note 7). The spatial structures and texture details are well preserved across all spectral bands, without obvious noise or structural artifacts observed. The synthesized RGB image generated from the reconstructed HSI cube exhibits high visual consistency with the image directly captured by a conventional color camera in both color distribution and overall appearance. In addition, four representative color regions were selected for reflectance spectrum analysis. Figure 4(c) compares the reference and reconstructed spectra, yielding an average RMSE of 0.03. These results demonstrate the capability of the proposed framework for stable and accurate spectral reconstruction in multi-color scenes.

**D. Hyperspectral Classification Applications**

To further validate the performance of the proposed method in practical spectral classification tasks, we conducted experiments on counterfeit fruit identification (Figure 5) and in vivo oral imaging and classification (Figure 6).

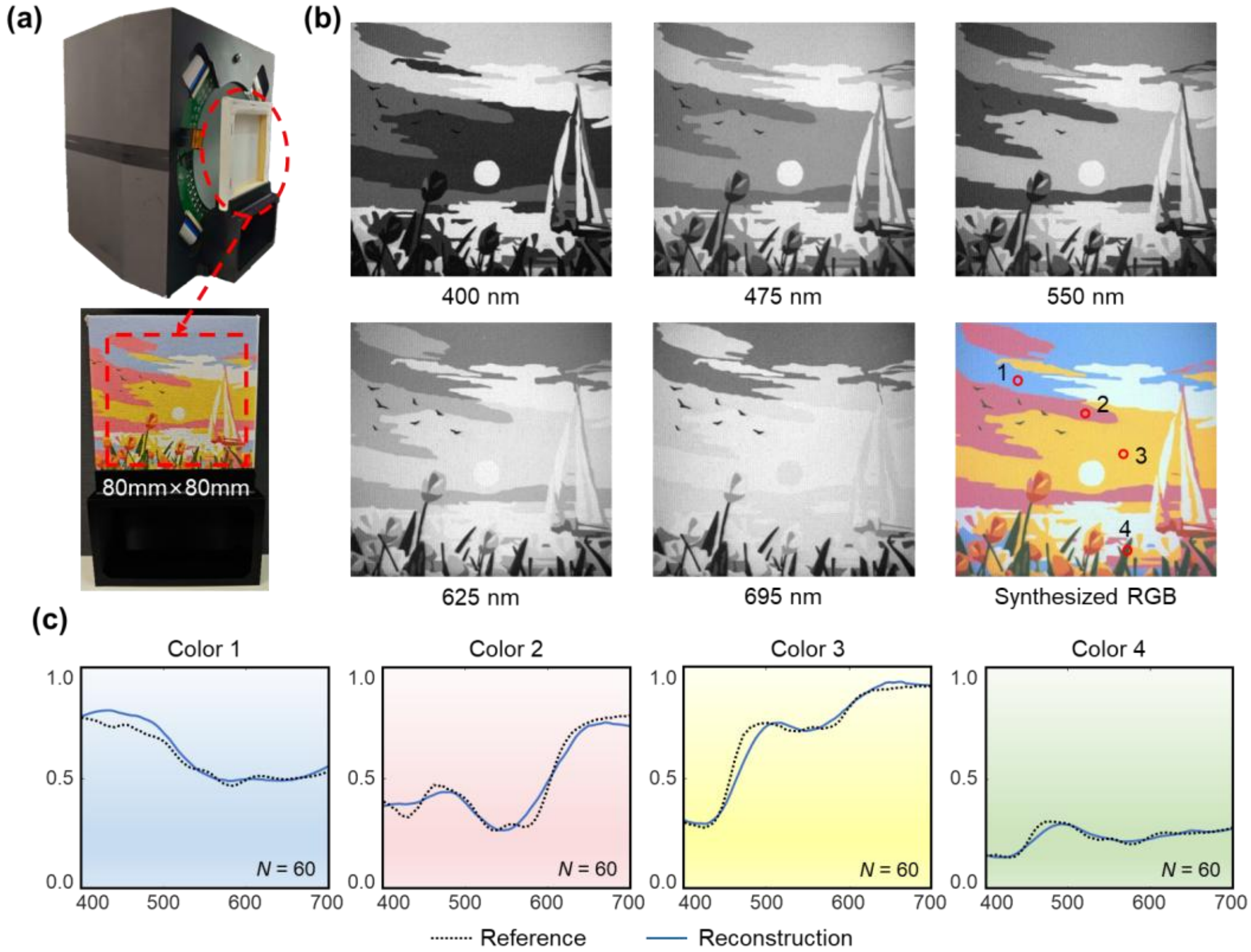

**Figure 4.** Hyperspectral imaging results of ROSE. (a) Schematic of the experimental setup with the reflective object. (b) Reconstructed images at 5 representative wavelengths and the synthesized RGB image. (c) Reconstructed spectral curves at 4 sampled locations, corresponding to the synthesized RGB image of (b).

Figure 5 presents the experimental results for distinguishing real and counterfeit lemons, demonstrating the practical utility of the proposed adaptive spectral-resolution framework. As shown in Figure 5(a), the target objects consist of a counterfeit lemon coated with yellow paint on the left and a real lemon on the right. Figure 5(b) presents the reconstructed hyperspectral data cube along with the corresponding synthesized RGB images (more results can be found in Supplementary Note 7).

Initially, hyperspectral data were acquired using the wavelength-encoded illumination patterns that optimized for the broadband range of 400–700 nm, reconstructing 60 spectral bands. The reconstructed spectra are presented in the top panel of Figure 5(c). By comparing the reconstructed and reference spectral curves, a pronounced discrepancy between the real and counterfeit lemons can be observed within the 400–550 nm range, whereas the spectral differences in the 550–700 nm region remain relatively minor. Based on this observation, the ROSE framework was employed to adaptively optimize the encoding strategy specifically for the targeted 400–550 nm band, reconstructing 60 spectral bands within this narrower region. The corresponding high-fidelity reconstructed spectra within this critical spectral region are shown in the bottom panel of Figure 5(c). Subsequently, a K-means clustering algorithm (*38*) based on the reconstructed spectral signatures was applied for target classification. The resulting classification maps are shown in Figure 5(d), together with the corresponding t-distributed stochastic neighbor embedding (t-SNE) feature visualizations (*39*) in Figure 5(e) to evaluate the spectra-based image segmentation.

Compared with the initial classification result with the illumination patterns that optimized for the broadband spectra (400–700 nm), the latter results obtained with the illumination patterns that optimized for the narrow band (400–550 nm) exhibit significantly improved spectral curve consistency and image segmentation accuracy. This enhancement is further validated by the t-SNE visualizations: after applying the task-specific illumination pattern optimization, the feature clusters corresponding to the real and counterfeit lemons become considerably more compact and distinctly separated in the embedded space. Quantitatively, the classification intersection over union (IoU) (*40*) increases substantially from 0.9253 to 0.9983. These findings confirm that task-specific optimization of illumination patterns toward feature-critical spectral bands can effectively improve both spectral consistency and segmentation accuracy in specific classification tasks.

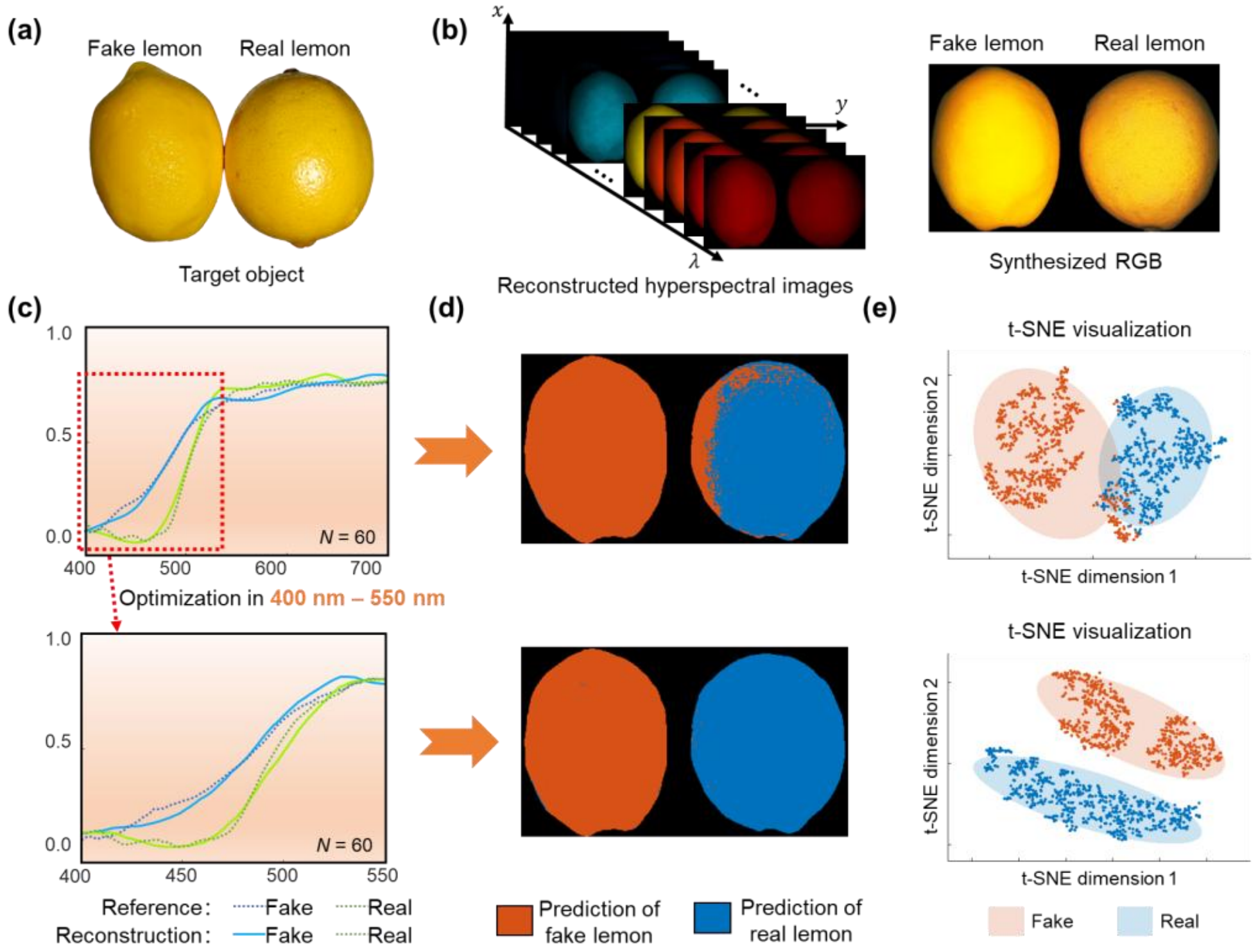

**Figure 5.** Experimental results for distinguishing real and fake lemons using the adaptive spectral resolution framework. (a) Target objects used in the experiment. (b) Visualization of the reconstructed hyperspectral data cube and synthesized RGB images. (c) Comparison of reconstructed and reference spectral curves using the 400–700 nm and optimized 400–550 nm narrowband encoding strategy. (d) K-Means clustering classification maps based on the broadband and optimized narrowband reconstructed spectra. (e) Corresponding t-SNE feature visualizations evaluating spectral separability before and after resolution optimization.

Figure 6 presents the hyperspectral oral imaging and classification results obtained using an oral diagnostic device developed based on ROSE. As shown in Figure 6(a), during data acquisition, the volunteers simply position their oral cavity at the imaging port of the device. Figure 6(b) presents the reconstructed hyperspectral data cube, while Figure 6(c) shows the synthesized RGB image of the volunteer's teeth and gingiva (more results can be found in Supplementary Note 7). Clinical diagnosis confirmed the presence of pigment deposition on the upper anterior teeth, as well as dental calculus accumulation near the gingival margin of the lower incisors.

To evaluate the diagnostic utility of the reconstructed spectral data, a K-Means clustering algorithm was employed to perform unsupervised classification of different tissue regions based on their spectral signatures. Figures 6(d) and 6(e) demonstrate the hyperspectral-based tissue classification results for the upper and lower dental regions, respectively. Figure 6(d) successfully distinguishes the regions with pigment deposition from normal teeth, while Figure 6(e) accurately segments the normal teeth, gingiva tissue, and dental calculus regions. Both the classification maps exhibit spatially continuous and well-defined segmentation regions, showing strong consistency with the clinically observed dental conditions. The corresponding t-SNE visualizations further corroborate the high spectral separability among these different tissue types. These experimental results demonstrate that the hyperspectral data acquired using the ROSE framework can effectively characterize subtle spectral differences among complex oral tissues, thereby enabling robust spectral-based automatic tissue classification.

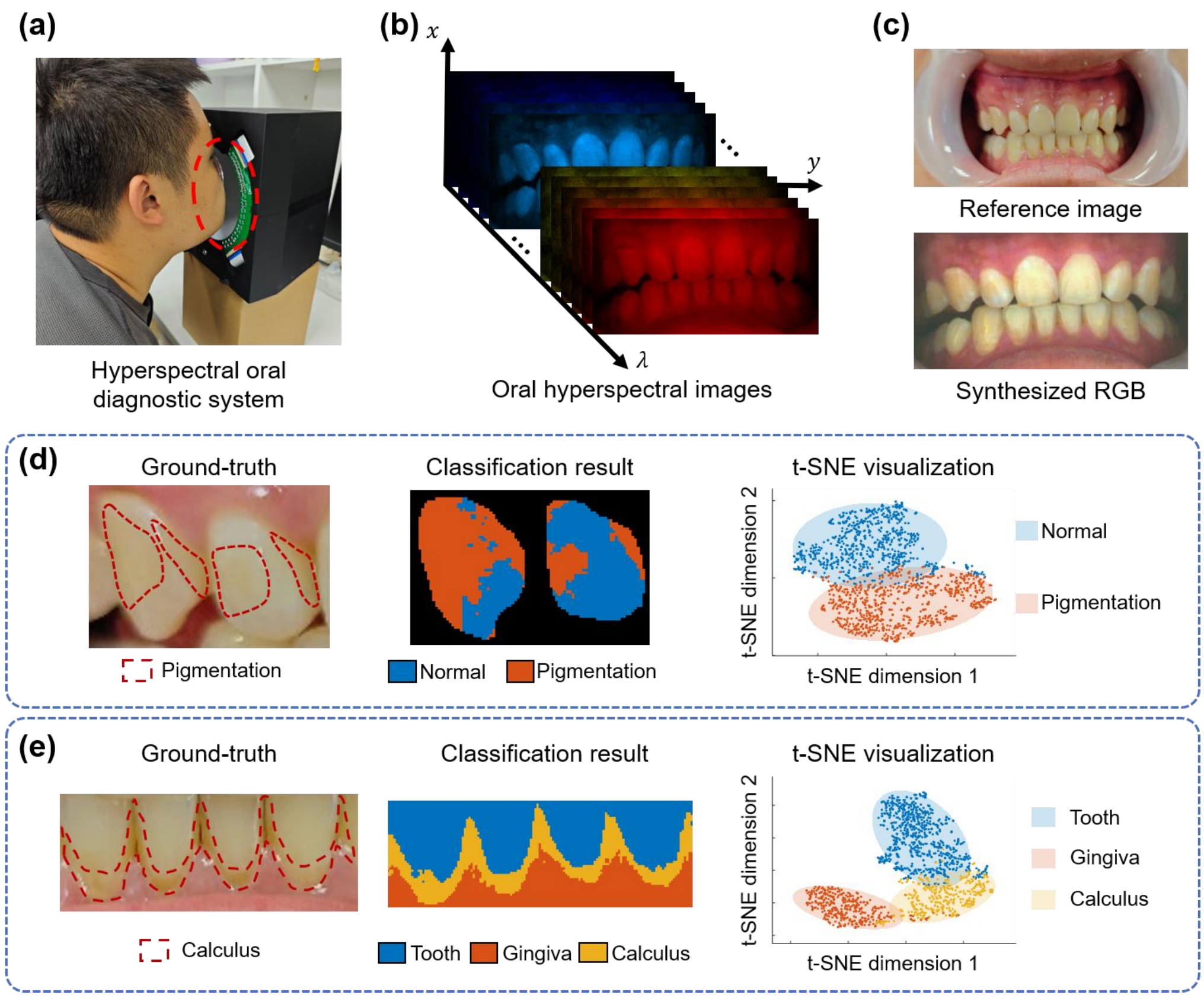

**Figure 6.** Hyperspectral imaging and classification results for oral health-assisted diagnosis. (a) Photograph of the hyperspectral oral diagnostic device developed based on the proposed method. (b) Reconstructed hyperspectral data cube. (c) Synthesized RGB image and the reference image of the volunteer's teeth and gingiva. (d) Spectral classification map of pigment deposition on the upper teeth and the corresponding t-SNE feature visualization. (e) Spectral classification map of normal teeth, gingiva, and dental calculus on the lower teeth, alongside the corresponding t-SNE feature visualization.

**E. Extension to Existing Optical Systems**

Since the wavelength encoding process is implemented entirely in the illumination path, the proposed illumination module can be readily integrated into existing optical imaging platforms without requiring modifications to the detection system. We demonstrate the integration capability of ROSE in both a fiber bundle-based imaging system and a microscopic imaging system, as shown in Fig. 7 (see more details in Supplementary Note 8 and Supplementary Note 9).

Figures 7(a)–7(c) present the schematic configuration and imaging results of the fiber-based HSI system. The encoded illumination is projected onto a USAF 1951 resolution test chart combined with distinct color filters. The transmitted light is coupled into a fiber bundle via a lens group, and the data are captured at the imaging end using a standard monochrome camera. As observed from the reconstructed spectral images in Fig. 7(b), the system can clearly resolve the line structures corresponding to Group-1, Elements 3 and 4 on the resolution chart. Moreover, the reconstructed spectral curves extracted from different filtered regions exhibit high agreement with the reference spectra (Fig. 7(c)).

Figures 7(d)–7(f) display the application results of the microscopic HSI system. As illustrated in Fig. 7(d), the illumination module is directly employed as the light source of a microscope system. Using a 10× objective, hyperspectral data acquisition was performed on a hematoxylin and eosin-stained ovarian tissue slice. The reconstruction results reveal that the system not only preserves the high-resolution tissue morphology and cellular structural details with high fidelity, but also accurately acquires the spectral signatures of various tissue positions. More results can be found in Supplementary Note 7.

These results demonstrate that ROSE can function as a plug-and-play stochastic illumination module, enabling conventional imaging platforms to achieve HSI through the simple replacement of the illumination source.

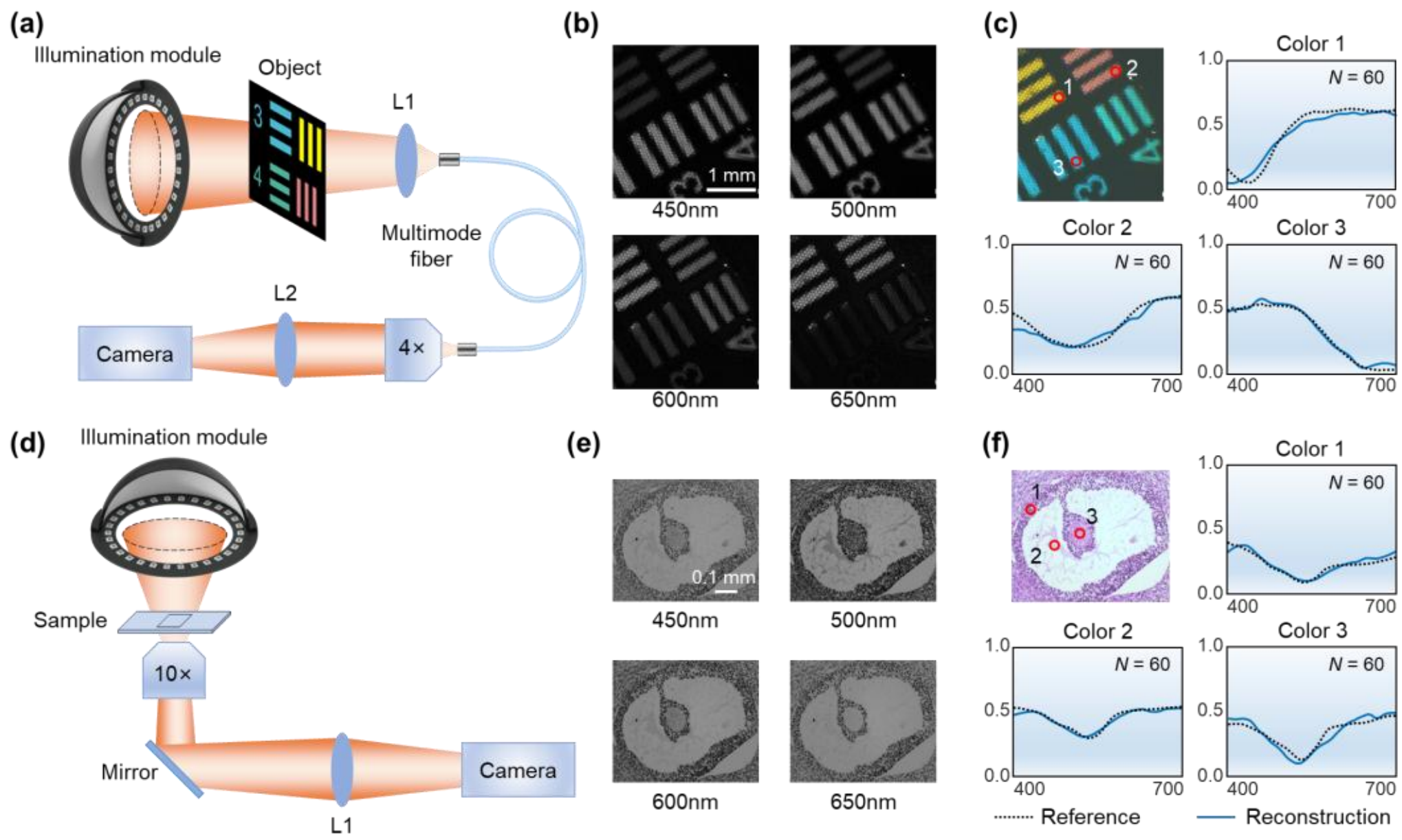

**Figure 7.** Extended applications of ROSE in fiber-based and microscopic imaging systems. (a) Schematic configuration of the fiber-based HSI system. (b) Reconstructed images of the USAF 1951 resolution test chart. (c) Comparison of reconstructed and reference spectral curves for different filtered regions on the resolution chart. (d) Schematic configuration of the microscopic HSI system. (e) Reconstructed images of the H&E-stained ovarian tissue slice. (f) Comparison of reconstructed and reference spectral curves for different tissue positions.

# CONCLUSIONS

In summary, we present the ROSE framework as a reconfigurable wavelength-encoded illumination module with stochastic spectra, for active, high-resolution HSI. By transferring the spectral encoding process from the optical detection path to the illumination side, the proposed method enables highly efficient and flexible hyperspectral image acquisition and reconstruction. This reconfigurable wavelength encoding is achieved through the synergistic combination of a programmable narrowband LED array and PWM-based intensity modulation. Meanwhile, the GA-based optimization framework is introduced to autonomously search for task-specific optimal illumination strategies. This strategy allows for significantly improved spectral reconstruction accuracy and the adaptive enhancement of spectral resolution in designated spectral bands.

Experimental results demonstrate that the proposed ROSE method achieves accurate and stable spectral reconstruction in both broadband and band-specific imaging tasks. Crucially, by leveraging the GA-based optimization framework to optimize wavelength encoding for specific targeted bands, the system is capable of capturing fine-grained spectral features that are often smoothed out by fixed encoding schemes. This adaptive enhancement of spectral resolution in feature-critical regions significantly improves the discriminative power of the reconstructed data, leading to a substantial boost in the accuracy of spectral classification tasks. Furthermore, because wavelength encoding is fully implemented at the illumination side, the system exhibits exceptional compatibility with existing imaging platforms. We have experimentally validated this plug-and-play capability by successfully integrating the ROSE module into both microscopic (*41, 42*) and endoscopic (*43, 44*) imaging systems. Finally, practical classification experiments, including oral health-assisted diagnosis, further validate its broad applicability. The proposed ROSE framework provides a versatile and scalable solution for adaptive HSI, opening new opportunities for intelligent spectral sensing across biomedical, industrial, and scientific imaging applications.

## METHODS

### A. System Design

The encoded illumination light source employs four cascaded TM1809 chip-level controllers. Each chip independently controls 9 LED channels, enabling the cascaded configuration to drive up to 36 narrowband LED channels. More detailed circuit design is described in Supplementary Note 2.

To mix and homogenize the light of different spectral bands emitted by the LEDs, the interior of the optical cavity is coated with a highly reflective barium sulfate diffuse reflection layer. An STM32 microcontroller uses TTL signals to synchronize the coded illumination sequences with the exposure of a monochrome CMOS camera (MER2-302-56U3M, Daheng Imaging, Beijing, China). To average out intensity fluctuations induced by PWM switching, the camera exposure time is set to 10 times the 400 Hz PWM scanning period of the controllers. This temporal integration yields a stable, effective acquisition frame rate of 40 fps.

### B. Encoded Illumination Optimization and Reconstruction

For system encoding optimization, an X-Rite ColorChecker Classic (24-color chart) is utilized as the reference target. The ground-truth reflectance spectra of these color patches are measured using a fiber spectrometer (EK2000-Pro, Choptics Instruments, Shanghai, China). To adaptively optimize the encoded illumination patterns for specific detection tasks, a genetic algorithm is employed. The genetic algorithm initializes with a population size of 20, where the chromosome encoding for each illumination pattern stores the selection indices of 4 out of the 30 LED channels alongside their corresponding intensity weights (more details of experiments can be found in Supplementary Note 4 and Supplementary Note 5). During the iterative optimization process, the crossover probability is set to 0.7, and a mutation operation is applied with a probability of 0.2. The mutation operation involves randomly replacing the selected LED indices and introducing Gaussian white noise with a standard deviation of 0.1 to the intensity weights, followed by normalization. The optimization process terminates after a maximum of 150 iterations. Following the acquisition of the wavelength-encoded measurements under the optimized illumination, the hyperspectral data cubes are recovered using the total variation augmented Lagrangian alternating direction algorithm (TVAL3) (*45*) (see more details in Supplementary Note 6).

## Acknowledgements

This work was supported by the National Natural Science Foundation of China (62573029, 62503032, 62222304).

## Data availability

All data supporting the findings of this study are available within the article and its Supplementary Information files and from the corresponding authors upon reasonable request.

## Additional information

Supplementary information: Please refer to the supporting documents.

## Code availability

All code will be provided upon request to the authors. Please address your requests for data and code to the corresponding authors.

## Competing interests

The authors declare no competing interests.